%% file: main.tex
\documentclass[conference]{IEEEtran}
\IEEEoverridecommandlockouts
\usepackage{subcaption}
\usepackage{cite}
\usepackage{amsmath,amssymb,amsfonts}
\usepackage{algorithmic}
\usepackage{graphicx}
\usepackage{textcomp}
\usepackage{xcolor}
\usepackage{balance}
\usepackage{booktabs}
\usepackage{multirow}
\def\BibTeX{{\rm B\kern-.05em{\sc i\kern-.025em b}\kern-.08em
    T\kern-.1667em\lower.7ex\hbox{E}\kern-.125emX}}
\begin{document}

\title{RAG-CSI: Global–Local Retrieval via Joint State and Dynamics Matching for CSI Prediction}

\author{\IEEEauthorblockN{Xinyue~Jiang$^{\dag}$, Tong~Li$^{\dag}$, Shiceng~Zhang$^{\S}$, Ke~Chen$^{\ddag}$, Zhu~Xiao$^{\dag}$} 
\IEEEauthorblockA{$^{\dag}$College of Computer Science and Electronic Engineering, Hunan University, Changsha, China.
\\$^{\S}$Xidian University, Xian, China.
\\$^{\ddag}$Peng Cheng Laboratory, Shenzhen 518066, China.
}}

\maketitle
\begin{abstract}
Channel state information (CSI) prediction forecasts future channel states from historical observations,
supporting transceiver design and resource allocation with reduced measurement overhead.
However, existing deep learning predictors mainly encode channel knowledge in model parameters,
which can limit their ability to represent diverse patterns of channel dynamics across scenes.
In this paper, we propose RAG-CSI, a retrieval-enhanced Transformer framework that augments prediction with relevant historical channel knowledge.
Specifically, we construct a global--local CSI knowledge store containing representative observations across scenes and historical observations from the current serving scene.
We further develop joint state--dynamics retrieval (JSDR), which jointly characterizes channel states, phase dynamics, and amplitude dynamics to retrieve relevant CSI sequences. A Transformer predictor then uses cross-attention to fuse the retrieved knowledge with the current CSI history and predict future CSI with limited additional model parameters. Experiments on a multi-scene CSI dataset show overall improvements exceeding 10\% across all five evaluation metrics over the best baseline for each metric, including a 15.42\% reduction in normalized mean squared error.
\end{abstract}

\begin{IEEEkeywords}
CSI prediction, Channel Knowledge, Retrieval-Augmented Generation (RAG), Transformer.
\end{IEEEkeywords}

\input{Intro}

\input{Method}
\input{Experiment}

\input{Related}

\input{Conclusion}

\bibliographystyle{IEEEtran}
\bibliography{ref}


\end{document}

%% file: Intro.tex
\section{Introduction}
Channel state information (CSI) describes the propagation characteristics of wireless links and provides the basis for transceiver design and resource allocation~\cite{guo2025prompt,guo2022overview,zhou20263dckm,jiang2026pihg}. 
CSI acquisition typically relies on pilot-based channel measurements and can incur substantial signaling overhead, 
motivating CSI prediction to reduce the need for additional measurements~\cite{sattari2026csi,zhou2024transformer}. 
CSI prediction typically forecasts future channel states from historical observations by exploiting temporal correlations in CSI. 
This task is challenging because multipath propagation and user mobility induce complex channel variations across time, antennas, and subcarriers.

Deep learning has been increasingly explored for CSI prediction owing to its ability to learn nonlinear dependencies from channel observations~\cite{gao2026ai}. 
Representative architectures include recurrent neural networks with long short-term memory (LSTM) or gated recurrent unit (GRU) modules~\cite{LSTM}, 
and Transformers that capture temporal dependencies through attention mechanisms~\cite{Transformer,luan2023channelformer,jin2025linformer}. More recently, wireless foundation models such as WiFo~\cite{WIFO} have extended prediction across different channel configurations through training on heterogeneous CSI datasets. Despite these advances, existing deep learning methods primarily rely on network parameters to encode channel patterns. As the scale and diversity of CSI data increase, limited model capacity can constrain the representation of channel dynamics patterns across various scenes.

\begin{figure}[tb]
    \centering
    \vspace{-2mm}
    \includegraphics[width=\linewidth]{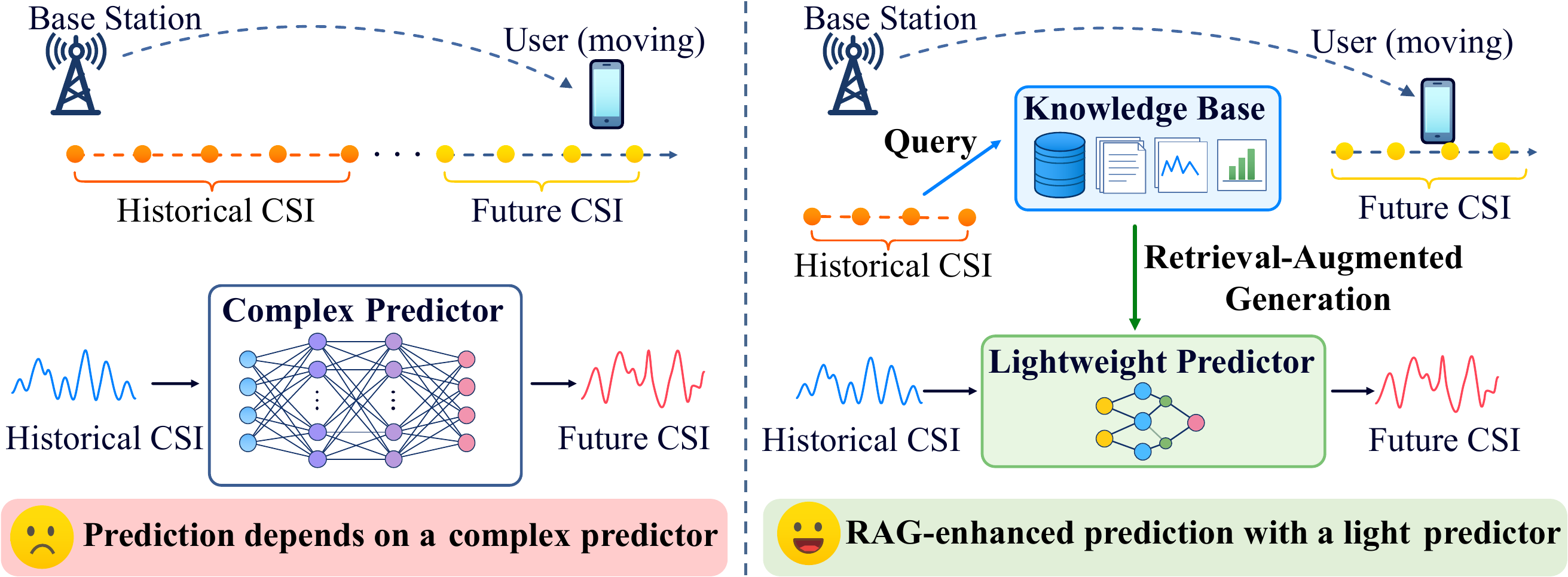}
    \vspace{-4mm}
    \caption{Comparison of conventional and retrieval-augmented CSI prediction frameworks.}
    \label{fig_intro}
    \vspace{-5mm}
\end{figure}

In this paper, we aim to develop a CSI prediction framework that accommodates diverse channel dynamics with a modest number of model parameters. Achieving such an objective involves three challenges. \textbf{First}, CSI prediction requires learning complex channel dynamics from high-dimensional historical observations, but limited observation windows and model capacity restrict the use of diverse channel dynamics patterns. \textbf{Second}, differences in propagation conditions across scenes can lead to distinct CSI statistics and temporal variation patterns. Consequently, learned global channel knowledge may not adequately support CSI prediction in a specific scene. \textbf{Third}, CSI exhibits complex variations across time, antennas, and subcarriers. Sequences with similar overall values may exhibit different directions of phase dynamics and different amplitude trends. Constructing a relevance measure to capture the channel dynamics patterns from limited historical observations is challenging.

To address these challenges, we propose RAG-CSI, a retrieval-enhanced framework for CSI prediction across diverse scenes. As illustrated in Fig.~\ref{fig_intro}, RAG-CSI retrieves relevant channel knowledge from external memory to augment the context available to the predictor. Specifically, we construct a global--local CSI knowledge store to provide complementary references. The global store, maintained on the user side, contains representative historical CSI accumulated or preloaded across scenes, while the local store, maintained at the serving base station, contains previously observed CSI from the current scene. We then develop a joint state--dynamics retrieval method to identify references relevant to the prediction task. Using historical complex-valued CSI, we construct descriptors of channel states, phase dynamics, and amplitude dynamics. Normalized complex conjugate products characterize phase dynamics, while logarithmic amplitude differences capture amplitude dynamics. These descriptors are independently normalized and combined with weights into a unified retrieval representation, enabling the retrieval of references with similar channel states and historical dynamics patterns to provide informative context for CSI prediction. Finally, a Transformer predictor integrates the retrieved channel knowledge with the input CSI history through cross-attention, enabling prediction to benefit from both learned patterns and explicitly retrieved references.

Our main contributions are summarized as follows:
\begin{itemize}
    \item We investigate retrieval-enhanced CSI prediction by explicitly exploiting historical channel observations beyond the current input window. The proposed framework supplements historical observations with relevant channel knowledge retrieved using CSI, expanding the available prediction context and reducing reliance on knowledge encoded solely in model parameters.

    \item We propose RAG-CSI, a retrieval-enhanced Transformer framework for CSI prediction. RAG-CSI combines global and local knowledge stores to provide complementary channel references, and develops joint state--dynamics retrieval (JSDR) to identify relevant references by jointly considering channel states, phase dynamics, and amplitude dynamics. The retrieved knowledge is integrated through cross-attention to improve prediction accuracy with limited additional model parameters.

    \item  We conduct experiments on a multi-scene CSI dataset to evaluate RAG-CSI. The results demonstrate that RAG-CSI outperforms state-of-the-art baselines by more than 10\% across all five metrics: mean squared error (MSE), normalized mean squared error (NMSE), mean absolute error (MAE), normalized mean absolute error (NMAE), and cosine similarity $\rho$.
\end{itemize}

%% file: Method.tex
\begin{figure*}[tb]
    \centering
    \includegraphics[width=0.85\linewidth]{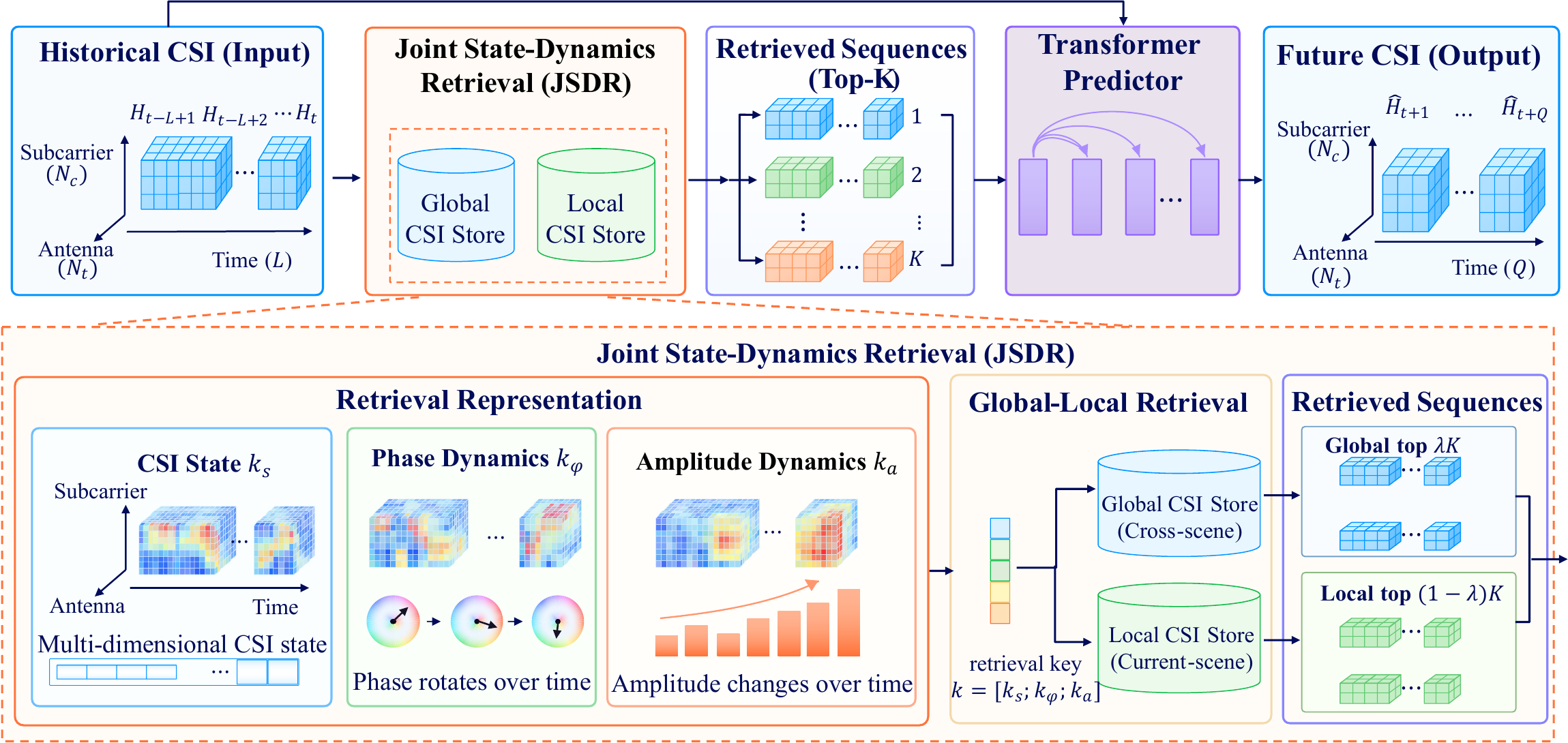}
    \vspace{-1mm}
    \caption{The overview of the proposed RAG-CSI. The framework consists of JDSR, a Transformer predictor, and CSI reconstruction. JSDR combines CSI states with phase and amplitude dynamics to retrieve relevant sequences from global and local CSI stores, offering historical references for Transformer-based prediction.}
    \label{fig_overview}
    \vspace{-4mm}
\end{figure*}

\section{System Model}
\label{sec2}

\subsection{Channel Model}
We consider a MIMO-OFDM system in which a base station (BS) equipped with $N_t$ antennas serves multiple single-antenna users over $N_c$ subcarriers. Although the BS serves multiple users within a propagation scene, channel prediction is performed on a per-user basis. For simplicity, we describe the channel between the BS and one user. The transmitted signal reaches the user through multiple propagation paths, and the resulting channel is represented as the superposition of $P$ paths. The channel vector on the $n$-th subcarrier is written as
\begin{equation}
    \mathbf{h}_n
    =
    \sum_{p=1}^{P}
    \alpha_p
    e^{-j2\pi f_n \tau_p}
    \mathbf{a}(f_n,\mathbf{s}_p),
    \label{eq:channel_vector}
\end{equation}
where $\alpha_p\in\mathbb{C}$, $\tau_p$, and $\mathbf{s}_p$ denote the complex path gain, propagation delay, and departure direction of the $p$-th path, respectively, and $f_n$ is the frequency of the $n$-th subcarrier. The array response vector $\mathbf{a}(f_n,\mathbf{s}_p)\in\mathbb{C}^{N_t}$ is given by
\begin{equation}
    \mathbf{a}(f_n,\mathbf{s}_p)
    =
    \left[
    e^{-j\frac{2\pi f_n}{c}\mathbf{r}_1^{T}\mathbf{s}_p},
    e^{-j\frac{2\pi f_n}{c}\mathbf{r}_2^{T}\mathbf{s}_p},
    \ldots,
    e^{-j\frac{2\pi f_n}{c}\mathbf{r}_{N_t}^{T}\mathbf{s}_p}
    \right]^T,
    \label{eq:array_response}
\end{equation}
where $c$ is the speed of light and $\mathbf{r}_m$ denotes the three-dimensional position vector of the $m$-th antenna relative to a reference antenna. For a uniform linear array (ULA), this response reduces to a phase progression determined by the antenna spacing and angle of departure. Stacking the channel vectors across all subcarriers yields the complete CSI matrix
\begin{equation}
    \mathbf{H}
    =
    \left[
    \mathbf{h}_1,\mathbf{h}_2,\ldots,\mathbf{h}_{N_c}
    \right]
    \in\mathbb{C}^{N_t\times N_c},
    \label{eq:csi_matrix}
\end{equation}
where the $n$-th column $\mathbf{h}_n\in\mathbb{C}^{N_t}$ represents the channel vector on the $n$-th subcarrier.

\subsection{Problem Definition}
CSI prediction estimates future channels from historical observations. We use DeepMIMO~\cite{DeepMIMO} to generate CSI at sampled user locations under specified antenna and OFDM configurations, and then organize the CSI snapshots along simulated user trajectories in temporal order. Let $\mathbf{H}_t\in\mathbb{C}^{N_t\times N_c}$ denote the CSI at time $t$. The historical and future CSI sequences are
\begin{equation}
\begin{aligned}
    \mathcal{H}_t^{\mathrm{hist}}
    &=\left(\mathbf{H}_{t-L+1},\ldots,\mathbf{H}_t\right),\\
    \mathcal{H}_t^{\mathrm{fut}}
    &=\left(\mathbf{H}_{t+1},\ldots,\mathbf{H}_{t+Q}\right),
\end{aligned}
\label{eq:csi_sequences}
\end{equation}
where $L$ and $Q$ are the history and prediction lengths, respectively, with
$\mathcal{H}_t^{\mathrm{hist}}\in\mathbb{C}^{L\times N_t\times N_c}$ and
$\mathcal{H}_t^{\mathrm{fut}}\in\mathbb{C}^{Q\times N_t\times N_c}$.

A conventional predictor estimates future CSI solely from the observed history:
\begin{equation}
    \widehat{\mathcal{H}}_t^{\mathrm{fut}}
    =F_{\theta}\!\left(\mathcal{H}_t^{\mathrm{hist}}\right).
    \label{eq:conventional_prediction}
\end{equation}
In RAG-CSI, relevant channel knowledge is additionally retrieved from a CSI memory $\mathcal{M}$:
\begin{equation}
\begin{aligned}
    \mathcal{K}_t
    &=\mathcal{R}\!\left(\mathcal{H}_t^{\mathrm{hist}};\mathcal{M}\right),\\
    \widehat{\mathcal{H}}_t^{\mathrm{fut}}
    &=F_{\theta}\!\left(\mathcal{H}_t^{\mathrm{hist}},\mathcal{K}_t\right),
\end{aligned}
\label{eq:retrieval_prediction}
\end{equation}
where $\mathcal{R}(\cdot)$ denotes retrieval and $\mathcal{K}_t$ denotes the retrieved channel knowledge. The predictor is trained by minimizing
\begin{equation}
    \min_{\theta}\;
    \mathcal{L}\!\left(
        F_{\theta}\!\left(
        \mathcal{H}_t^{\mathrm{hist}},\mathcal{K}_t
        \right),
        \mathcal{H}_t^{\mathrm{fut}}
    \right),
    \label{eq:prediction_objective}
\end{equation}
where $\mathcal{L}(\cdot,\cdot)$ measures the prediction error.

\section{Method}

\subsection{Overview}
RAG-CSI is a retrieval-augmented framework that enhances Transformer-based CSI prediction with relevant historical channel knowledge. As shown in Fig.~\ref{fig_overview}, it consists of a Global--Local CSI knowledge store, Joint State--Dynamics Retrieval (JSDR), and a Transformer predictor with cross-attention fusion. The Global store contains representative CSI across scenes, while the Local store maintains historical CSI from the current serving scene. Given the observed CSI history, JSDR jointly considers channel states, phase dynamics, and amplitude dynamics to retrieve relevant sequences from both stores. Cross-attention then fuses the retrieved knowledge with the historical representation, and the Transformer encoder--decoder models channel dynamics to predict future CSI through a projection layer.

\subsection{Retrieval-Augmented Generation (RAG) for CSI Prediction}
Retrieval-augmented generation (RAG) enhances a parametric model with relevant information retrieved from external memory. For CSI prediction, it introduces reusable historical channel patterns beyond the current observation window. As shown in Fig.~\ref{fig_overview}, RAG-CSI uses $\mathcal{H}_t^{\mathrm{hist}}$ as the query to retrieve channel references  from the knowledge store $\mathcal{M}$, which are then incorporated into the Transformer to predict CSI.

\subsubsection{Global--Local CSI Knowledge Store}
To capture both cross-scene and local channel knowledge, we construct
$\mathcal{M}=(\mathcal{M}_{\mathrm{G}},\mathcal{M}_{\mathrm{L}})$.
The Global store $\mathcal{M}_{\mathrm{G}}$ is maintained on the user side and contains historical CSI knowledge accumulated or preloaded across multiple scenes, while the Local store $\mathcal{M}_{\mathrm{L}}$ is maintained at the serving BS and contains knowledge collected within its serving area. Both stores use the same storage format, enabling the current CSI history $\mathcal{H}_t^{\mathrm{hist}}$ to retrieve complementary references from both sources.

\subsubsection{Global--Local CSI Top-$K$ Retrieval}
Given $\mathcal{H}_t^{\mathrm{hist}}$, a total of $K$ reference windows are retrieved from the two stores. A balancing factor $\lambda\in[0,1]$ determines the nominal allocation:
\begin{equation}
    K_{\mathrm{G}}=\operatorname{round}(\lambda K),
    \qquad
    K_{\mathrm{L}}=K-K_{\mathrm{G}}.
    \label{eq:retrieval_allocation}
\end{equation}
If the Local store contains fewer than $K_{\mathrm{L}}$ entries, all available Local entries are used and the remaining budget is assigned to the Global store. Retrieval from each store is independently performed using the proposed JSDR similarity:
\begin{equation}
    \mathcal{K}_t^{a}
    =
    \operatorname{TopK}_{K_a}\!\left(
    \mathcal{M}_a;\,
    S_{\mathrm{JSDR}}\!\left(
    \mathcal{H}_t^{\mathrm{hist}},\cdot
    \right)
    \right),
    \quad a\in\{\mathrm{G},\mathrm{L}\},
    \label{eq:global_local_retrieval}
\end{equation}
where $\operatorname{TopK}_{K_a}$ selects the $K_a$ entries with the highest similarity scores. The retrieved knowledge is represented as
\begin{equation}
    \mathcal{K}_t
    =
    \left(
    \mathcal{K}_t^{\mathrm{G}},
    \mathcal{K}_t^{\mathrm{L}}
    \right)
    =
    \mathcal{R}\!\left(
    \mathcal{H}_t^{\mathrm{hist}};\mathcal{M}
    \right).
    \label{eq:retrieved_knowledge}
\end{equation}

\subsection{Joint State--Dynamics Retrieval (JSDR)}
Existing retrieval-augmented forecasting methods commonly use cosine similarity or Euclidean distance to retrieve knowledge~\cite{RAST,ma2026quizsf}. However, MIMO-OFDM CSI varies complexly across time, antennas, and subcarriers, similarity measures applied directly to CSI observations may not explicitly distinguish channel states from phase and amplitude dynamics. To address this, we propose JSDR, which characterizes channel states from historical CSI, phase dynamics via normalized complex conjugate products, and amplitude dynamics via logarithmic magnitude differences. These descriptors are normalized and combined into a unified retrieval key for more informative CSI knowledge retrieval.


\subsubsection{CSI State Representation}
We represent historical CSI using temporally weighted real and imaginary components. Each observation is vectorized as
\begin{equation}
    \mathbf{x}_{t,\ell}
    =
    \begin{bmatrix}
        \operatorname{vec}\!\left(\operatorname{Re}(\mathbf{H}_{t-L+\ell})\right) \\
        \operatorname{vec}\!\left(\operatorname{Im}(\mathbf{H}_{t-L+\ell})\right)
    \end{bmatrix},
    \quad \ell=1,\ldots,L,
    \label{eq:csi_state_observation}
\end{equation}
where $\operatorname{vec}(\cdot)$ follows a fixed antenna--subcarrier order. With temporal weights $r_\ell=\gamma^{L-\ell}$ and $0<\gamma\leq1$, we concatenate the weighted observations chronologically:
\begin{equation}
    \mathbf{k}_{\mathrm{s}}\!\left(\mathcal{H}_t^{\mathrm{hist}}\right)
    =
    \left[
        \sqrt{r_1}\mathbf{x}_{t,1}^{T},
        \ldots,
        \sqrt{r_L}\mathbf{x}_{t,L}^{T}
    \right]^{T}
    .
    \label{eq:weighted_csi_state}
\end{equation}
The descriptor is subsequently $\ell_2$-normalized. Smaller $\gamma$ emphasizes recent observations, whereas $\gamma=1$ assigns uniform weights. The same representation is applied to query and stored CSI sequences.

\subsubsection{Phase-Dynamics Representation}
To identify CSI knowledge with similar phase dynamics while avoiding phase-wrapping discontinuities, we use the normalized complex conjugate product at temporal lag $\ell$:
\begin{equation}
    \mathbf{G}_t^{(\ell)}
    =
    \frac{
        \mathbf{H}_t\mathbf{H}_{t-\ell}^{*}
    }{
        |\mathbf{H}_t||\mathbf{H}_{t-\ell}|
    },
    \label{eq:phase_change}
\end{equation}
where $(\cdot)^*$ denotes complex conjugation and all operations are element-wise. For nonzero coefficients, this product retains relative phase changes while removing magnitudes. The phase-dynamics descriptor is obtained through $f_{\phi}(\cdot)$:
\begin{equation}
    \mathbf{k}_{\phi}\!\left(
        \mathcal{H}_t^{\mathrm{hist}}
    \right)
    =
    f_{\phi}\!\left(
        \left\{\mathbf{G}_{u}^{(\ell)}\right\}
    \right),
    \label{eq:phase_dynamics_descriptor}
\end{equation}
where the collection includes valid observation pairs at the selected lags within $\mathcal{H}_t^{\mathrm{hist}}$, capturing phase dynamics over different intervals. Consistent with the CSI state representation, $f_{\phi}(\cdot)$ vectorizes the real and imaginary components, applies temporal weighting to emphasize recent changes, and concatenates the features in a fixed temporal, lag, and antenna--subcarrier order, followed by $\ell_2$ normalization.

\subsubsection{Amplitude-Dynamics Representation}
We characterize relative amplitude dynamics using logarithmic magnitude differences between consecutive observations:
\begin{equation}
    \mathbf{D}_t
    =
    \log|\mathbf{H}_t|
    -
    \log|\mathbf{H}_{t-1}|,
    \label{eq:amplitude_change}
\end{equation}
where all operations are element-wise. This representation captures relative amplitude changes while reducing sensitivity to absolute channel magnitudes.

The amplitude-dynamics descriptor is
\begin{equation}
    \mathbf{k}_a\!\left(
        \mathcal{H}_t^{\mathrm{hist}}
    \right)
    =
    f_a\!\left(
        \left\{\mathbf{D}_u\right\}
    \right),
    \label{eq:amplitude_dynamics_descriptor}
\end{equation}
where $f_a(\cdot)$ applies temporal weighting, vectorizes and concatenates the amplitude changes in temporal and antenna--subcarrier order, and performs $\ell_2$ normalization.



\subsubsection{Joint State--Dynamics Retrieval Key}
We combine the CSI state, phase-dynamics, and amplitude-dynamics descriptors into a unified retrieval key:
\begin{equation}
    \mathbf{k}\!\left(\mathcal{H}_t^{\mathrm{hist}}\right)
    =
    \operatorname{L2Norm}\!\left(
    \left[
        \sqrt{\omega_s}\mathbf{k}_s;
        \sqrt{\omega_\phi}\mathbf{k}_\phi;
        \sqrt{\omega_a}\mathbf{k}_a
    \right]
    \right),
    \label{eq:jsdr_key}
\end{equation}
where $\omega_s,\omega_\phi,\omega_a\geq0$ sum to one and control the contributions of the three descriptors. For a stored CSI sequence $\mathcal{H}_i^{\mathrm{ref}}$, the retrieval similarity is computed as
\begin{equation}
    S_{\mathrm{JSDR}}\!\left(
        \mathcal{H}_t^{\mathrm{hist}},
        \mathcal{H}_i^{\mathrm{ref}}
    \right)
    =
    \mathbf{k}\!\left(\mathcal{H}_t^{\mathrm{hist}}\right)^{T}
    \mathbf{k}\!\left(\mathcal{H}_i^{\mathrm{ref}}\right).
    \label{eq:jsdr_similarity}
\end{equation}
This similarity guides Global--Local Top-$K$ retrieval based on joint channel-state and dynamics information.

\subsection{Retrieval-Augmented Prediction}

We use a lightweight Transformer backbone for CSI prediction.
Cross-attention fuses features of the current CSI history with
representations of the retrieved sequences, producing a
retrieval-enhanced historical representation. The backbone
models temporal dependencies from this enriched context and
generates features for future channel states. A linear projection
followed by reshaping produces the real and imaginary components
of the predicted CSI sequence
$\widehat{\mathcal{H}}_t^{\mathrm{fut}}
\in\mathbb{C}^{Q\times N_t\times N_c}$.

%% file: Experiment.tex
\section{Experiments}

\subsection{Experimental settings}
We evaluate RAG-CSI on a multi-scene CSI dataset generated using the open-source DeepMIMO framework~\cite{DeepMIMO}. DeepMIMO provides ray-tracing-based channel data from realistic wireless environments, supporting the evaluation of CSI prediction across different propagation scenes. Using DeepMIMO v4 scenario configurations, we generate MIMO-OFDM channel samples from 10 scenes, including ASU Campus, New York, Miami, and San Diego. Within each scene, we simulate user trajectories and assemble the corresponding channel observations into CSI sequences. The historical sequence length is set to 12, and the prediction horizon is one step. For retrieval, the default number of retrieved CSI sequences is set to $K=10$, and the Global--Local balancing factor is set to $\lambda=0.5$. The generated sequences are divided into training and test sets at a ratio of 8:2. The detailed dataset properties and channel settings are summarized in Table~\ref{tab_dataset}.

\begin{table}[tb]
    \centering
    \vspace*{4pt}
    \caption{Dataset properties and experimental settings.}
    \label{tab_dataset}
    \small
    \setlength{\tabcolsep}{3pt}
    \begin{tabular}{c|c}
        \toprule
        \textbf{Properties} & \textbf{Value} \\
        \midrule
        Number of Scenes & 10 \\
        Number of antennas & 8 \\
        Antenna array form & ULA \\
        Number of subcarriers & 8 \\
        Number of paths & 25 \\
        Carrier frequency & 3.5 GHz \\
        Bandwidth & 20 MHz \\
        Historical CSI sequence length & 12 \\
        Prediction horizon & 1 \\
        \bottomrule
    \end{tabular}
    \vspace{-4mm}
\end{table}

\subsection{Baselines and Evaluation Metrics}
We compare RAG-CSI with six representative baselines:

\textbf{LSTM~\cite{LSTM}:} A recurrent predictor that uses LSTM layers to learn temporal dependencies from historical CSI for future channel prediction.

\textbf{LSTM-GRU~\cite{LSTM-GRU}:} A hybrid architecture comprising cascaded LSTM and GRU blocks for CSI prediction. 

\textbf{Transformer~\cite{Transformer}:} A transformer-based prediction framework for parallel time-domain channel prediction.

\textbf{RCDNet~\cite{RCDNet}:} A recurrent channel-deduction network that uses channel mixing to model spatial-frequency correlations and an LSTM to capture temporal correlations.

\textbf{WiFo~\cite{WIFO}:} A wireless foundation model pretrained through masked CSI reconstruction across space, time, and frequency, and applied to channel prediction.

\textbf{PromptFM~\cite{PromptFM}:} A prompt-guided foundation model designed for CSI modeling across multiple frequency bands and evaluated here on CSI prediction.

We evaluate prediction performance using mean squared error (MSE), normalized mean squared error (NMSE), mean absolute error (MAE), normalized mean absolute error (NMAE), and cosine correlation $\rho$~\cite{RCDNet}. Lower values of the four error metrics and a higher $\rho$ indicate better prediction performance.

\subsection{Overall Performance}
Table~\ref{tab_overall_performance} summarizes the overall CSI prediction performance. RAG-CSI achieves the best results across all five metrics. Compared with the best baseline for each metric, RAG-CSI reduces MSE, NMSE, MAE, and NMAE by 14.36\%, 15.42\%, 10.56\%, and 10.13\%, respectively, while improving cosine correlation $\rho$ by 11.72\%. The reductions in both absolute and normalized errors indicate more accurate CSI prediction, while the higher $\rho$ reflects better alignment between the predicted and ground-truth channel vectors. These results support the effectiveness of augmenting historical observations with relevant channel knowledge. 

\begin{table}[t]
    \centering
    \vspace*{4pt}
    \caption{Overall CSI prediction performance comparison.}
    \label{tab_overall_performance}
    \small
    \setlength{\tabcolsep}{3pt}
    \begin{tabular}{lccccc}
        \toprule
        \textbf{Method}
        & \textbf{MSE}$\downarrow$
        & \textbf{NMSE}$\downarrow$
        & \textbf{MAE}$\downarrow$
        & \textbf{NMAE}$\downarrow$
        & $\rho\uparrow$ \\
        \midrule
        LSTM~\cite{LSTM}
        & 0.8301 & 0.7395 & 0.6883 & 0.7863
        & \underline{0.7115} \\

        LSTM-GRU~\cite{LSTM-GRU}
        & 1.1752 & 1.0031 & 0.8915 & 1.0013 & 0.3614 \\

        Transformer~\cite{Transformer}
        & 0.8150 & \underline{0.7296}
        & \underline{0.6765} & \underline{0.7756} & 0.6776 \\

        RCDNet~\cite{RCDNet}
        & \underline{0.8128} & 0.7396 & 0.6773 & 0.7793 & 0.6982 \\

        WiFO~\cite{WIFO}
        & 1.0039 & 0.8727 & 0.7941 & 0.8996 & 0.5047 \\
        PromptFM~\cite{PromptFM}
        & 0.9850 & 0.8602 & 0.7785 & 0.8812 & 0.4131 \\
        \textbf{RAG-CSI}
        & \textbf{0.6961} & \textbf{0.6172}
        & \textbf{0.6051} & \textbf{0.6970} & \textbf{0.7949} \\
        \midrule
        \textbf{Improv.}
        & \textbf{14.36\%} & \textbf{15.42\%}
        & \textbf{10.56\%} & \textbf{10.13\%} & \textbf{11.72\%} \\
        \bottomrule
    \end{tabular}
    \vspace{-5mm}
\end{table}

\subsection{Performance Across Scenes}
Table~\ref{tab_scene_performance} compares CSI prediction across four different scenes. RAG-CSI consistently achieves the lowest NMSE and NMAE and the highest $\rho$ across all four scenes. Compared with the best baseline for each scene and metric, it reduces NMSE by 4.17\%--12.74\% and NMAE by 4.93\%--7.76\%, while improving $\rho$ by 4.06\%--12.06\%, with the largest gains observed in Miami. These results demonstrate the effectiveness of RAG-CSI across diverse propagation environments.

\begin{table*}[t]
    \centering
    \small
    \caption{CSI prediction performance across different scenes.}
    \vspace{-2mm}
    \label{tab_scene_performance}
    \resizebox{\linewidth}{!}{%
    \begin{tabular}{l*{12}{c}}
        \toprule
        \textbf{Scenes}
        & \multicolumn{3}{c}{\textbf{ASU Campus}}
        & \multicolumn{3}{c}{\textbf{New York}}
        & \multicolumn{3}{c}{\textbf{Miami}}
        & \multicolumn{3}{c}{\textbf{San Diego}} \\
        
        \cmidrule(lr){2-4}
        \cmidrule(lr){5-7}
        \cmidrule(lr){8-10}
        \cmidrule(lr){11-13}
        \textbf{Methods}
        & NMSE$\downarrow$ & NMAE$\downarrow$ & $\rho\uparrow$
        & NMSE$\downarrow$ & NMAE$\downarrow$ & $\rho\uparrow$
        & NMSE$\downarrow$ & NMAE$\downarrow$ & $\rho\uparrow$
        & NMSE$\downarrow$ & NMAE$\downarrow$ & $\rho\uparrow$ \\
        \midrule

        LSTM~\cite{LSTM}
        & \underline{0.6528} & \underline{0.7112} & \underline{0.7904}
        & \underline{0.6939} & 0.7921 & \underline{0.7299}
        & 1.0133 & 1.0054 & 0.4542
        & \underline{0.6038} & \underline{0.7186} & \underline{0.7240} \\

        LSTM-GRU~\cite{LSTM-GRU}
        & 1.0043 & 1.0020 & 0.2954
        & 0.9995 & 0.9999 & 0.3053
        & 1.0008 & 0.9989 & \underline{0.6889}
        & 1.0006 & 1.0005 & 0.3005 \\

        Transformer~\cite{Transformer}
        & 0.6745 & 0.7339 & 0.7202
        & 0.7778 & 0.8283 & 0.6573
        & \underline{0.7586} & \underline{0.7800} & 0.6684
        & 0.6466 & 0.7492 & 0.6493 \\

        RCDNet~\cite{RCDNet}
        & 0.7012 & 0.7462 & 0.7260
        & 0.7018 & \underline{0.7826} & 0.6887
        & 0.8110 & 0.8394 & 0.6750
        & 0.6558 & 0.7490 & 0.6939 \\

        WiFO~\cite{WIFO}
        & 0.8435 & 0.8713 & 0.5763
        & 0.8486 & 0.9016 & 0.4397
        & 1.0035 & 1.0016 & 0.2779
        & 0.7508 & 0.8319 & 0.6177 \\

        PromptFM~\cite{PromptFM}
        & 0.8277 & 0.8494 & 0.4547
        & 0.8163 & 0.8852 & 0.4716
        & 1.0031 & 1.0005 & 0.0857
        & 0.6718 & 0.7749 & 0.6620 \\
        \textbf{RAG-CSI}
        & \textbf{0.5819} & \textbf{0.6717} & \textbf{0.8225}
        & \textbf{0.6650} & \textbf{0.7440} & \textbf{0.8100}
        & \textbf{0.6620} & \textbf{0.7195} & \textbf{0.7719}
        & \textbf{0.5570} & \textbf{0.6825} & \textbf{0.7677} \\
        \midrule
        \textbf{Improv.}
        & \textbf{10.86\%} & \textbf{5.55\%} & \textbf{4.06\%}
        & \textbf{4.17\%} & \textbf{4.93\%} & \textbf{10.97\%}
        & \textbf{12.74\%} & \textbf{7.76\%} & \textbf{12.06\%}
        & \textbf{7.75\%} & \textbf{5.02\%} & \textbf{6.03\%} \\
        \bottomrule
    \end{tabular}%
    }
    
\end{table*}

\subsection{Ablation Study}
We ablate retrieval augmentation and JSDR using two variants: \textit{w/o RAG}, which removes CSI sequence retrieval and uses only the Transformer-based backbone for prediction, and \textit{w/o JSDR}, which replaces JSDR with cosine correlation for CSI retrieval. As shown in Table~\ref{tab_ablation_result}, removing RAG increases NMSE from 0.6172 to 0.7344 and reduces $\rho$ from 0.7949 to 0.6351, confirming the benefit of retrieved CSI knowledge. Replacing JSDR also degrades performance, with NMSE increasing to 0.6783 and $\rho$ decreasing to 0.6844, demonstrating the importance of incorporating channel states together with phase and amplitude dynamics.

\begin{table}[tb]
  \vspace{-3mm}
    \centering
    \caption{Ablation study of RAG-CSI.}
    \vspace{-2mm}
    \label{tab_ablation_result}
    \resizebox{\linewidth}{!}{%
    \begin{tabular}{lccccc}
        \toprule
        \textbf{Method}
        & \textbf{MSE}$\downarrow$
        & \textbf{NMSE}$\downarrow$
        & \textbf{MAE}$\downarrow$
        & \textbf{NMAE}$\downarrow$
        & $\rho\uparrow$ \\
        \midrule

        w/o JSDR
        & \underline{0.7910}
        & \underline{0.6783}
        & \underline{0.6648}
        & \underline{0.7563}
        & \underline{0.6844} \\

        w/o RAG
        & 0.8388
        & 0.7344
        & 0.6970
        & 0.7943
        & 0.6351 \\

        \textbf{RAG-CSI}
        & \textbf{0.6961}
        & \textbf{0.6172}
        & \textbf{0.6051}
        & \textbf{0.6970}
        & \textbf{0.7949} \\
        \bottomrule
    \end{tabular}%
    }
    \vspace{-2mm}
\end{table}

\subsection{Hyperparameter Sensitivity}

\begin{figure}[tb]
    \centering
    \subfloat[]{\includegraphics[height=0.39\linewidth,trim=0 0 0 -7.2bp,clip]{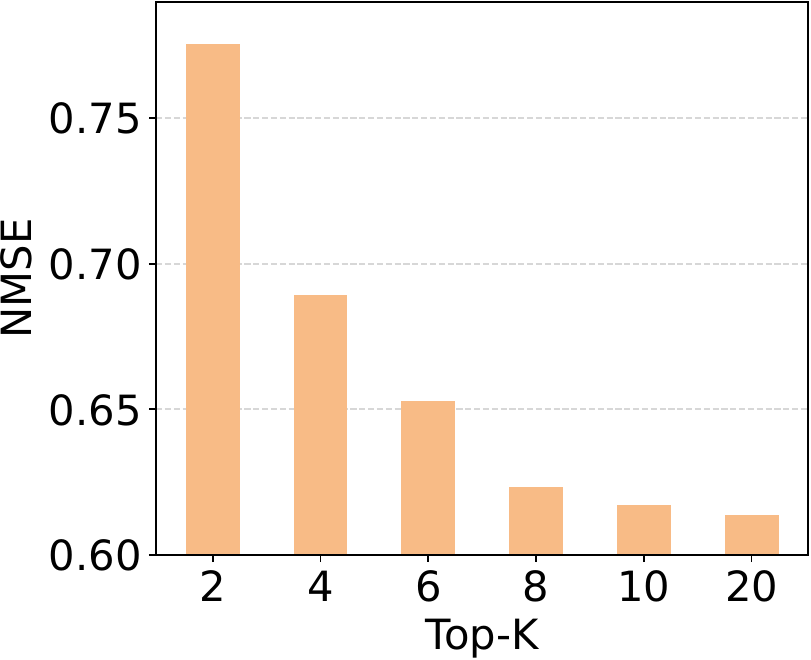} \label{topk_nmse}}
    \subfloat[]{\includegraphics[height=0.39\linewidth,trim=0 -4bp 0 0,clip]{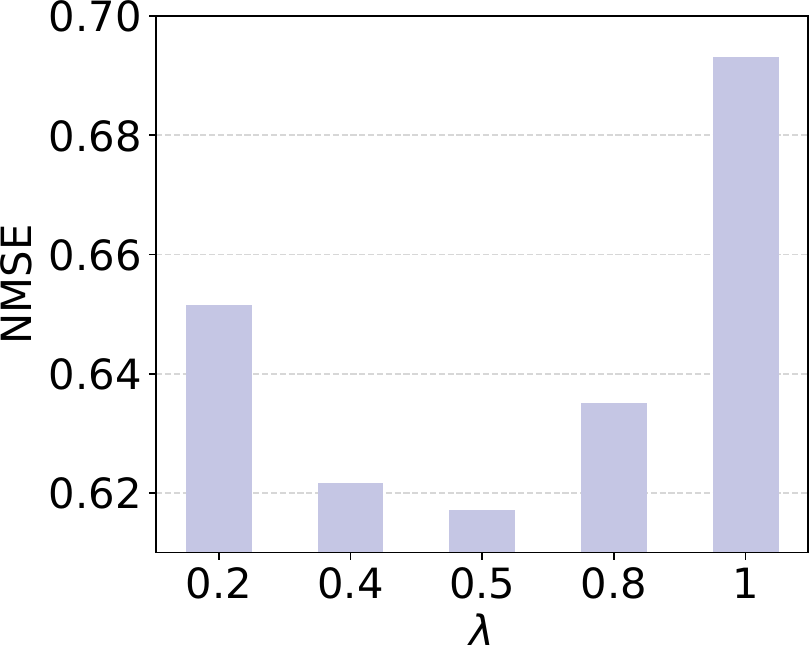}\label{lambda_nmse}}
    \vspace{-2mm}
    \caption{Hyperparameter sensitivity: (a) retrieval size $K$ with $\lambda=0.5$; (b) Global--Local balancing factor $\lambda$ with $K=10$.}
     \label{topk_performance}
    \vspace{-4mm}
\end{figure}

We evaluate the sensitivity of RAG-CSI to the retrieval size $K$ and the Global--Local balancing factor $\lambda$.
Fig.~\ref{topk_performance}(a) shows the results for $K\in\{2,4,6,8,10,20\}$ with $\lambda=0.5$. NMSE decreases markedly as $K$ increases to 8, while further increasing $K$ brings only marginal gains.
This indicates that a moderate number of retrieved sequences captures most of the benefit. Fig.~\ref{topk_performance}(b) shows the results for varying $\lambda$ with $K=10$. The best performance is achieved at $\lambda=0.5$, while shifting the allocation toward either Global or Local knowledge increases the error. These results support jointly using Global and Local knowledge, balancing access to diverse channel patterns with relevance to the current scene.

\subsection{Complexity Analysis}


Table~\ref{tab_complexity} compares parameter counts and average inference times per CSI sequence. RAG-CSI has the fewest parameters (0.451 million), 50.39\% fewer than the Transformer baseline. Its average inference time is 0.3242 ms per sequence, 65.72\% lower than that of the Transformer baseline and the third shortest overall, behind only LSTM and LSTM-GRU. Together with the accuracy gains above, these results demonstrate that RAG-CSI combines a compact prediction network with competitive inference efficiency.

\begin{table}[tb]
    \centering
    \caption{Model size and inference time comparison.}
    \label{tab_complexity}
    \vspace{-2mm}
    \small
    \setlength{\tabcolsep}{5pt}
    \begin{tabular}{lcc}
        \toprule
        \textbf{Method} & \textbf{Params (M)} &
        \textbf{Inference Time (ms)} \\
        \midrule
        LSTM~\cite{LSTM}             & 6.435 & 0.1422 \\
        LSTM-GRU~\cite{LSTM-GRU}      & 1.744 & 0.1718 \\
        Transformer~\cite{Transformer} & 0.909 &0.9457 \\
        RCDNet~\cite{RCDNet}          & 4.341 & 0.3791 \\
        WiFO~\cite{WIFO}              & 21.326 & 0.4114 \\
        PromptFM~\cite{PromptFM}      & 21.598 &  0.5042\\
        \textbf{RAG-CSI}                 & 0.451 & 0.3242  \\
        \bottomrule
    \end{tabular}
    \vspace{-4mm}
\end{table}

%% file: Related.tex
\section{Related Work}
CSI prediction exploits temporal correlations to mitigate channel aging and reduce pilot overhead.
Early methods used linear prediction~\cite{ARPrediction} or combined autoregression with principal component analysis~\cite{PCAPrediction}, but predefined models may struggle with complex channel dynamics.
Deep learning enables nonlinear CSI modeling~\cite{jiang2025unirm}.
LSTM and GRU were explored for fading-channel prediction~\cite{LSTM}, while Transformers support parallel multi-step prediction~\cite{Transformer}.
More recently, WiFo~\cite{WIFO} used masked CSI reconstruction for prediction across channel configurations.
However, these methods primarily encode channel knowledge in fixed model parameters, increasing demands on model capacity.

Retrieval-Augmented Generation (RAG) complements parametric models with knowledge retrieved from external memory, enabling access to information beyond model parameters. Initially developed for knowledge-intensive NLP tasks~\cite{RAG}, RAG has recently been extended to spatiotemporal forecasting~\cite{RAST}, where retrieved historical patterns provide additional prediction context. However, retrieval augmentation remains largely unexplored for CSI prediction, motivating the use of historical CSI knowledge to capture diverse channel dynamics without relying solely on model parameters.

%% file: Conclusion.tex
\section{Conclusion}
This paper proposed RAG-CSI, a retrieval-enhanced Transformer framework for CSI prediction across diverse propagation scenes. RAG-CSI combines a Global--Local CSI knowledge store, JSDR-based retrieval, and cross-attention fusion to exploit relevant historical channel knowledge beyond the current observation window. JSDR jointly captures channel states, phase dynamics, and amplitude dynamics to retrieve informative CSI sequences. Experiments on a multi-scene CSI dataset showed improvements exceeding 10\% over the best baselines across all five evaluation metrics, demonstrating the effectiveness of retrieval augmentation for CSI prediction.

%% file: ref.bib
@article{LSTM,
  title={Deep learning for fading channel prediction},
  author={Jiang, Wei and Schotten, Hans Dieter},
  journal={IEEE Open Journal of the Communications Society},
  volume={1},
  pages={320--332},
  year={2020},
  publisher={IEEE}
}

@article{Transformer,
  title={Accurate channel prediction based on transformer: Making mobility negligible},
  author={Jiang, Hao and Cui, Mingyao and Ng, Derrick Wing Kwan and Dai, Linglong},
  journal={IEEE Journal on Selected Areas in Communications},
  volume={40},
  number={9},
  pages={2717--2732},
  year={2022},
  publisher={IEEE}
}

@article{LSTM-GRU,
  title={{LSTM-GRU} model-based channel prediction for one-bit massive MIMO system},
  author={Helmy, Islam and Tarafder, Pulok and Choi, Wooyeol},
  journal={IEEE Transactions on Vehicular Technology},
  volume={72},
  number={8},
  pages={11053--11057},
  year={2023},
  publisher={IEEE}
}

@article{WIFO,
  title={WiFo: Wireless foundation model for channel prediction},
  author={Liu, Boxun and Gao, Shijian and Liu, Xuanyu and Cheng, Xiang and Yang, Liuqing},
  journal={Science China Information Sciences},
  volume={68},
  number={6},
  pages={162302},
  year={2025},
  publisher={Springer}
}

@article{RCDNet,
  title={Channel deduction: A new learning framework to acquire channel from outdated samples and coarse estimate},
  author={Chen, Zirui and Zhang, Zhaoyang and Yang, Zhaohui and Huang, Chongwen and Debbah, Merouane},
  journal={IEEE Journal on Selected Areas in Communications},
  volume={43},
  number={3},
  pages={944--958},
  year={2025},
  publisher={IEEE}
}

@inproceedings{ARPrediction,
  author = {Baddour, Kareem E. and Beaulieu, Norman C.},
  title = {Improved Pilot-Assisted Prediction of Unknown Time-Selective Rayleigh Channels},
  booktitle = {Proceedings of the IEEE International Conference on Communications (ICC)},
  volume = {11},
  pages = {5192--5199},
  year = {2006},
  doi = {10.1109/ICC.2006.255490}
}

@inproceedings{PCAPrediction,
  author = {Nagashima, Rei and Ohtsuki, Tomoaki and Jiang, Wenjie and Takatori, Yasushi and Nakagawa, Tadao},
  title = {Channel Prediction for Massive {MIMO} with Channel Compression Based on Principal Component Analysis},
  booktitle = {Proceedings of the IEEE 27th Annual International Symposium on Personal, Indoor, and Mobile Radio Communications (PIMRC)},
  pages = {1--6},
  year = {2016},
  doi = {10.1109/PIMRC.2016.7794949}
}

@inproceedings{RAG,
  author = {Lewis, Patrick and Perez, Ethan and Piktus, Aleksandra and Petroni, Fabio and Karpukhin, Vladimir and Goyal, Naman and K{\"u}ttler, Heinrich and Lewis, Mike and Yih, Wen-tau and Rockt{\"a}schel, Tim and Riedel, Sebastian and Kiela, Douwe},
  title = {Retrieval-Augmented Generation for Knowledge-Intensive {NLP} Tasks},
  booktitle = {Advances in Neural Information Processing Systems},
  volume = {33},
  pages = {9459--9474},
  year = {2020}
}

@inproceedings{RAST,
  author = {Ruan, Weilin and Dang, Xilin and Zhou, Ziyu and Lyu, Sisuo and Liang, Yuxuan},
  title = {A Retrieval Augmented Spatio-Temporal Framework for Traffic Prediction},
  booktitle = {Proceedings of the AAAI Conference on Artificial Intelligence},
  volume = {40},
  number = {46},
  pages = {39163--39172},
  year = {2026},
  doi = {10.1609/aaai.v40i46.41264}
}

@inproceedings{DeepMIMO,
  author = {Alkhateeb, Ahmed},
  title = {{DeepMIMO}: A Generic Deep Learning Dataset for Millimeter Wave and Massive {MIMO} Applications},
  booktitle = {Proceedings of the Information Theory and Applications Workshop (ITA)},
  pages = {1--8},
  year = {2019},
  url = {https://arxiv.org/abs/1902.06435}
}

@inproceedings{ma2026quizsf,
  title={{QuiZSF}: A Retrieval-Augmented Framework for Zero-Shot Time Series Forecasting},
  author={Ma, Shichao and Zhou, Zhengyang and Huang, Qihe and Wang, Binwu and Wang, Yang},
  booktitle={Proceedings of the ACM Web Conference 2026},
  pages={1947--1958},
  year={2026}
}

@article{gao2026ai,
  title={{AI}-driven channel state information {(CSI)} extrapolation for {6G}: Current situations, challenges and future research},
  author={Gao, Yuan and Lu, Zichen and Wu, Xinyi and Yu, Wenjun and Liu, Shengli and Du, Jianbo and Jin, Yanliang and Zhang, Shunqing and Chu, Xiaoli and Xu, Shugong},
  journal={IEEE communications surveys \& tutorials},
  year={2026},
  publisher={IEEE}
}

@article{guo2025prompt,
  title={Prompt-enabled large AI models for {CSI} feedback},
  author={Guo, Jiajia and Cui, Yiming and Wen, Chao-Kai and Jin, Shi},
  journal={IEEE Journal on Selected Areas in Communications},
  year={2025},
  publisher={IEEE}
}

@article{guo2022overview,
  title={Overview of deep learning-based {CSI} feedback in massive {MIMO} systems},
  author={Guo, Jiajia and Wen, Chao-Kai and Jin, Shi and Li, Geoffrey Ye},
  journal={IEEE Transactions on Communications},
  volume={70},
  number={12},
  pages={8017--8045},
  year={2022},
  publisher={IEEE}
}

@article{luan2023channelformer,
  title={Channelformer: Attention based neural solution for wireless channel estimation and effective online training},
  author={Luan, Dianxin and Thompson, John S},
  journal={IEEE Transactions on Wireless Communications},
  volume={22},
  number={10},
  pages={6562--6577},
  year={2023},
  publisher={IEEE}
}

@article{zhou2024transformer,
  title={Transformer network based channel prediction for CSI feedback enhancement in AI-native air interface},
  author={Zhou, Tao and Liu, Xiangping and Xiang, Zuowei and Zhang, Haitong and Ai, Bo and Liu, Liu and Jing, Xiaorong},
  journal={IEEE Transactions on Wireless Communications},
  volume={23},
  number={9},
  pages={11154--11167},
  year={2024},
  publisher={IEEE}
}

@article{jin2025linformer,
  title={{LinFormer}: A linear-based lightweight transformer architecture for time-aware MIMO channel prediction},
  author={Jin, Yanliang and Wu, Yifan and Gao, Yuan and Zhang, Shunqing and Xu, Shugong and Wang, Cheng-Xiang},
  journal={IEEE Transactions on Wireless Communications},
  volume={24},
  number={9},
  pages={7177--7190},
  year={2025},
  publisher={IEEE}
}

@article{sattari2026csi,
  title={{CSI} prediction using diffusion models},
  author={Sattari, Mehdi and Aliakbari, Javad and i Amat, Alexandre Graell and Svensson, Tommy},
  journal={IEEE Transactions on Wireless Communications},
  volume={25},
  pages={21418--21432},
  year={2026},
  publisher={IEEE}
}

@inproceedings{PromptFM,
  title={A Universal Prompt-Guided Foundation Model for Multi-band Channel State Information Modeling},
  author={Huang, Zihan and Li, Tong and Lv, Xingzai and Rui, Hua and Li, Yong},
  booktitle={ICC 2026-IEEE International Conference on Communications},
  pages={1--6},
  year={2026},
  organization={IEEE}
}

@article{jiang2026pihg,
  title={PIHG: Physics-Inspired Heterogeneous Graph Neural Networks for Multi-band Radio Map Prediction},
  author={Jiang, Xinyue and Li, Tong and Xiao, Zhu and Chen, Ke and Tai, Andy Chi Lok and Tang, Zhuo and Li, Kenli},
  journal={IEEE Transactions on Wireless Communications},
  year={2026},
  publisher={IEEE}
}

@article{jiang2025unirm,
  title={UniRM: A universal large model for multiband 3D radio map construction},
  author={Jiang, Xinyue and Li, Tong and Xiao, Zhu and Chen, Ke and Ma, Shuai and Wang, Zhaocheng and Li, Keqin},
  journal={IEEE Journal on Selected Areas in Communications},
  year={2025},
  publisher={IEEE}
}

@inproceedings{zhou20263dckm,
  title={3DCKM: A Large-Scale Multiband 3D Channel Knowledge Map Dataset for Low-Altitude Corridors},
  author={Zhou, Chenyang and Jiang, Xinyue and Li, Tong and Xiao, Zhu and Zhang, Jie},
  booktitle={2026 IEEE International Conference on Communications Workshops (ICC Workshops)},
  pages={1--6},
  year={2026},
  organization={IEEE}
}
